\documentclass[aps,pra,preprint,groupedaddress]{revtex4-1}

\usepackage{graphicx}
\usepackage{dcolumn}
\usepackage{bm}
\usepackage{color}
\usepackage{amsfonts,amsmath,amsthm,graphicx}

\usepackage{epstopdf}

\begin{document}

\title{Influence of chirping the Raman lasers in an atom gravimeter: phase shifts due to the Raman light shift and to the finite speed of light}

\author{B. Cheng, P. Gillot, S. Merlet, F. Pereira Dos Santos}     
\email[]{franck.pereira@obspm.fr}
\homepage[]{https://syrte.obspm.fr/spip/}

\affiliation{LNE-SYRTE, Observatoire de Paris, PSL Research University, CNRS, Sorbonne Universit\'es, UPMC Univ. Paris 06, 61 avenue de l'Observatoire, 75014 Paris, France}

\date{\today}

\begin{abstract} 
We present here an analysis of the influence of the frequency dependence of the Raman laser light shifts on the phase of a Raman-type atom gravimeter.  Frequency chirps are applied to the Raman lasers in order to compensate gravity and ensure the resonance of the Raman pulses during the interferometer.  We show that the change in the Raman light shift when this chirp is applied only to one of the two Raman lasers is enough to bias the gravity measurement by a fraction of $\mu$Gal\footnote{$1~\mu$Gal~=~$10^{-8}$~m/s$^2$}. We also show that this effect is not compensated when averaging over the two directions of the Raman wavevector $k$. This thus constitutes a limit to the rejection efficiency of the $k$-reversal technique. Our analysis allows us to separate this effect from the effect of the finite speed of light, which we find in perfect agreement with expected values. This study highlights the benefit of chirping symmetrically the two Raman lasers.

\end{abstract}

\pacs{}

\maketitle

\section{Introduction}

Inertial sensors based on atom interferometry~\cite{Borde1989, Kasevich1991,  Riehle1991, Canuel2006, McGuirk2002, Sorrentino2014} now compete with the state-of-the-art classical instruments, both in terms of sensitivity and accuracy \cite{Gustavson2000, Peters2001,Louchet-Chauvet2011, Hu2013,   Gillot2014}.  Yet, systematic effects in these atom interferometers are still the subject of investigation and their ultimate level of performance in terms of accuracy is still to be met.  Out of the many systematics that affect the phase of such interferometers, many are usually rejected by the $k$-reversal technique, \emph{i.e.}~alternating measurements using two opposite directions for the Raman effective wavevector $\vec{k}_\mathrm{eff}$.  This is possible because, unlike the gravity phase shift, these systematics do not depend on the direction of   $\vec{k}_\mathrm{eff}$~\cite{Peters2001,Louchet-Chauvet2011}. This is the case of the Raman one-photon light shift \cite{Weiss1994}, the quadratic Zeeman effect and the frequency dependent phase shifts in the electronics   hardware operating in radio frequency domain. The efficiency of this rejection is in principle limited by the difference between the trajectories of the atoms between these two interferometer configurations, due to the change in the direction of the momentum kick imparted to the atoms by the lasers. Nevertheless, this difference is in practice small with respect to the size of the trajectory, typically of order of a mm over about ten centimeters. The efficiency of this rejection was studied in \cite{Mehlstaubler2007}.

The one-photon Raman light shift is the differential light shift that the out-of-resonance Raman lasers imprint onto the two hyperfine states in the Raman-type interferometers.  This differential light shift can in most cases be canceled by adjusting the ratio between the two Raman lasers. If not, its effect is in principle canceled by the symmetry of the interferometer, which, thanks to the use of a $\pi$ pulse that exchanges
internal state at the middle of the interferometer, has no sensitivity to constant frequency shifts.  However, because the atoms in free fall expand in the finite-size Raman beams during the interferometer, the light shift seen by the atoms at the three pulses vary, leading to a residual parasitic phase shift. This effect is hopefully rejected as explained above by using the $k$-reversal technique.

The aim of this paper is to put the efficiency of this rejection into question. In particular, we show that such atom gravimeters can be biased by a
time-dependent light shift arising from the frequency chirp applied onto the Raman lasers to compensate for the Doppler shift. Recently, \cite{Zhan2015} has shown that the frequency change related to such a frequency chirp induces a modification of the power ratio between the two Raman lasers after their amplification in a common tampered amplifier, and thus of the differential light shift. Here, we find that compensating this Doppler shift by chirping only one of the two Raman lasers induces a change in the light shift, even for a constant intensity ratio between the lasers, that leads to a bias in the gravity measurement.

The frequency chirp here plays a similar role as the effect of the two photon light shift (TPLS) studied in \cite{Gauguet2008}: it makes the resonance condition of the Raman transition vary over the duration of the interferometer. However, the effect described in the present paper differs from the TPLS in its origin, as it arrises from a time dependence of the one photon light shift. In particular, this effect would still be present in the case where the Raman beams would be produced using a single pair of lasers, (one laser propagating downwards, the second upwards), a configuration in which the TPLS is absent. Instead, in the configuration we use here, the counterpropagating lasers are produced by retroreflecting two initially copropagating Raman lasers. This standard configuration, which among other advantages guarantees a better stability of the phase difference between the Raman lasers, leads indeed to the presence of two pairs of Raman lasers onto the atoms, one being non-resonant and responsible for the TPLS. Though the two effects add up here, they do not have the same scaling with the laser parameters (such as the Raman detunings and intensities etc.), so that one can evaluate independently their relative contribution to the phase of the interferometer.

In the following sections, we start by calculating the amplitude of the effect of the frequency chirp(s) of the laser onto the one photon light shift, and derive the amplitude of the resulting bias as a function of the Raman detuning. We then perform measurements of the interferometer phase, and thus of this bias, as a function of the frequency chirps applied to the lasers for different Raman detunings. We find a good agreement with the calculations. As a side product, the study performed here allows us to measure the effect of the finite speed of light~\cite{Kuroda1991} onto the interferometer as a function of the chirps applied to the Raman lasers, and confirms the analysis of \cite{Peters2001}.

\section{Raman light shift}\label{section2}

We start by recalling the general expression of the one-photon light shift onto the two hyperfine state $|f\rangle$ and $|e\rangle$ of an alkali atom in the presence of two Raman laser beams $R1$ and $R2$ detuned from the D2 line, respectively by $\Delta_{R1}$ and $\Delta_{R2}$ (not necessarily equal)
\begin{eqnarray}
 \Omega_f^{AC}&=&\sum_k{\frac{|\Omega_{k,f1}|^2}{4(\Delta_{R1}+\Delta_k)}+\frac{|\Omega_{k,f2}|^2}{4(\Delta_{R2}+\Delta_k-\Delta_{HFS})}},\\
 \Omega_e^{AC}&=&\sum_k{\frac{|\Omega_{k,e1}|^2}{4(\Delta_{R1}+\Delta_k+\Delta_{HFS})}+\frac{|\Omega_{k,e2}|^2}{4(\Delta_{R2}+\Delta_k)}},
 \label{eq:ls1}
\end{eqnarray}
where $\Omega_{k,mn}$ is the Rabi frequency of laser Raman $Rn$ between the ground state $|m\rangle$ and the excited state $|F'=k\rangle$, $\Delta_{HFS}$ is the angular frequency of the ground state hyperfine transition and $\Delta_k$ is the angular frequency of the hyperfine splittings in the excited state $|F'=k\rangle$.

\begin{figure}[h]
       \centering
              \includegraphics[width=8 cm]{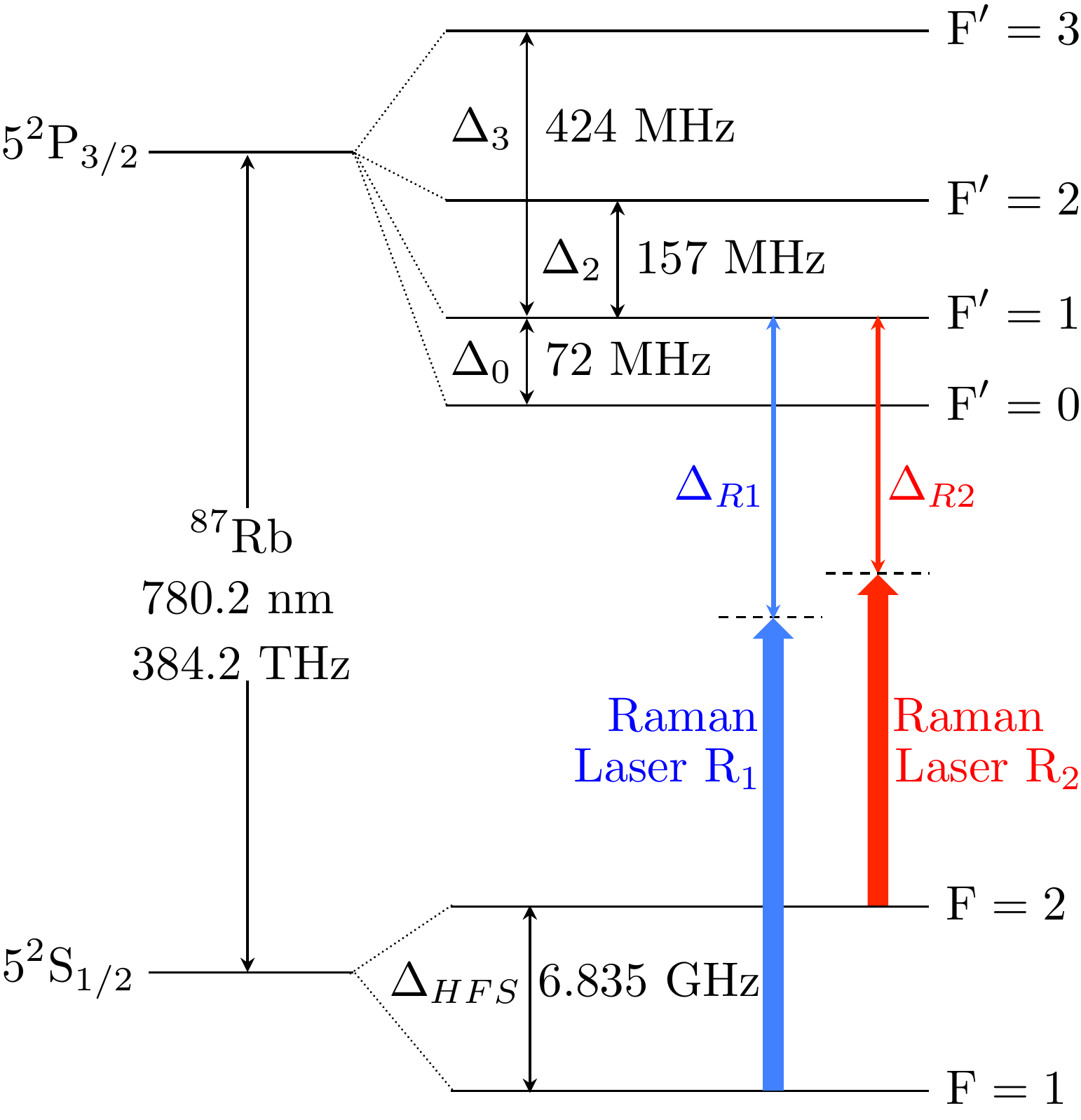}
\caption{(Color online) Relevant energy levels for $^{87}$Rb. $\Delta_{Ri}$ is   the Raman detuning for laser $Ri$, $\Delta_{i}$ is the frequency difference   between the $|F'=1\rangle$ level and the other $|F'\rangle$ levels of the $5^2P_{3/2}$ state.}
        \label{fig:levels}
\end{figure}

In the case of $^{87}$Rb atoms in the presence of two such Raman lasers, with the same circular polarization, these Rabi frequencies are given by

\begin{equation}\label{eq:lsf}
\Omega_f^{AC}=\frac{\Omega_1^2}{4}(\frac{5}{24\Delta_{R1}}+\frac{1}{8(\Delta_{R1}-\Delta_2)})+\frac{\Omega_2^2}{4}(\frac{5}{24(\Delta_{R1}-\Delta_{HFS})}+\frac{1}{8(\Delta_{R1}-\Delta_2-\Delta_{HFS})}),
\end{equation}
\begin{equation}\label{eq:lse}
\begin{split}
\Omega_e^{AC}&=\frac{\Omega_1^2}{4}(\frac{1}{120(\Delta_{R1}+\Delta_{HFS})}+\frac{1}{8(\Delta_{R1}-\Delta_2+\Delta_{HFS})}+\frac{1}{5(\Delta_{R1}-\Delta_3+\Delta_{HFS})})\\
 & \quad +\frac{\Omega^2}{4}(\frac{1}{120\Delta_{R2}}+\frac{1}{8(\Delta_{R2}-\Delta_2)}+\frac{1}{5(\Delta_{R2}-\Delta_3)}),
\end{split}
\end{equation}
where $\Omega_i$ are the simplified Rabi frequency $\Omega_i=DE_i/\hbar$, $D$ is the magnitude of the dipole moment of the D2 transition ($D=3.58\times10^{-29}~$C.m), and $E_i$ the amplitude of the electric field of the laser $Ri$. Here $\Delta_{Ri}$ and
$\Delta_i$ are defined taking the level $|F'=1\rangle$ as a reference (Figure~\ref{fig:levels}). Note that for the present paper, the
detunings $\Delta_{Ri}$ are to be considered in the frame of the atoms. For an atom at a velocity $v$, this detuning $\Delta_{Ri}$ is thus Doppler shifted by $\vec{k}_i.\vec{v}$ with respect to its corresponding atomic transition (the $|F=i\rangle \rightarrow |F'=1\rangle$ transition), \emph{i.e.}, $\Delta_{Ri}=\Delta_{Ri0}-\vec{k}_i.\vec{v}$, where $\Delta_{Ri0}$ is the detuning of the Raman laser $R1$ in the laboratory frame (or equivalently for an atom at rest).

\subsection{Condition for driving Raman transitions}

The two lasers $R1$ and $R2$ are used to induce stimulated Raman transitions between the two hyperfine states. The resonance condition in the laboratory frame for this process is given by:
$\omega_{R1}-\omega_{R2}=\omega_{HFS}+\omega_{Doppler}+\omega_{rec}$, where $\omega_{R1}$ and $\omega_{R2}$ are the laser frequencies in the laboratory frame, and $\omega_{Doppler} = (\vec{k}_1 - \vec{k}_2).\vec{v}$ and $\omega_{rec}= \hbar \mid \vec{k}_1 - \vec{k}_2 \mid^2/(2M)$ are the Doppler and recoil terms respectively~\cite{Moler1992}. This is equivalent to
\begin{equation}
\Delta_{R1}-\Delta_{R2} = \omega_{rec}.
\label{eq:condres}
\end{equation}

As for the coupling, it is characterized by the two-photon Rabi frequency, given by
\begin{equation}
\Omega_\mathrm{eff}=\frac{\Omega_1\Omega_2}{2}(\frac{1}{24\Delta}+\frac{1}{8(\Delta-\Delta_2)}),
\label{eq:rabifreq}
\end{equation}
where $\Delta$ is the almost common Raman detuning $\Delta \sim \Delta_{R1} \sim \Delta_{R2}$, as $\omega_{rec}= 2 \pi \times 15~$kHz~$<<\Delta_{Ri}$ (typically on the order of 1~GHz).

\subsection{Effect on the gravity induced chirp on the Differential Light Shift}

For atoms in free fall in the vacuum chamber, the Doppler shift increases linearly with time. In our experiment, we typically let the frequency of the laser $R1$ fixed, so that the detuning $\Delta_{R1}$ varies as $\Delta_{R1}= \Delta_{R10}-\vec{k}_1.\vec{g}t$. To fulfill the resonance
condition (eq. \ref{eq:condres}), we deliberately apply a frequency ramp onto the second Raman laser (that would otherwise be Doppler shifted in the opposite direction, as being counterpropagating). This way, the frequency of both lasers in the atomic frame change by the same amount $-\vec{k}_1.\vec{g}t$.

We calculate the effect of this frequency change on the Differential Light Shift (DLS) for the following typical parameters: $Delta_{R10}$~=~-0.923~GHz, and the Raman laser intensities of $I_{R1}$~=~11~mW/cm$^2$, $I_{R2}$~=~1.74~$\times$~$I_{R1}$. We take into account that in our configuration where the Raman lasers are retro-reflected on a common reference mirror, the light shift is doubled. The corresponding Rabi frequency $\Omega_\mathrm{eff}$ is $2\pi$~$\times$~22.7~kHz.

Using equations \ref{eq:lsf} and \ref{eq:lse}, we calculate a change in the light shift of $\Delta \delta_{DLS}~=~10.5~$Hz between the first and the last pulse of the interferometer, separated by a total interferometer time of 2$T$~=~160~ms. This induces a phase shift onto the interferometer of $\Delta\Phi$~=~$\Delta\delta_{DLS}/\Omega_\mathrm{eff}$~=~-0.42~mrad, and a corresponding gravity shift of -0.41~$\mu$Gal.

\subsection{Discussion on the influence of the experimental parameters}

Remarkably, in the configuration described above where the frequency chirp is applied to only one of the Raman lasers, this phase shift cannot be separated from the gravity induced phase shift using the usual technique of reversing the direction of the Raman wavevector. Indeed, this reversal requires to change the sign of the frequency chirp. The chirp induced light shift thus changes sign, as well as the gravity phase shift, so that these two contributions cannot be separated by averaging over these two measurements. Also, with contrast to the TPLS, the phase shift does not depend on the Rabi frequency: choosing a different Raman power changes in the same proportion the light shift and the Rabi frequency.

\begin{figure}
        \includegraphics{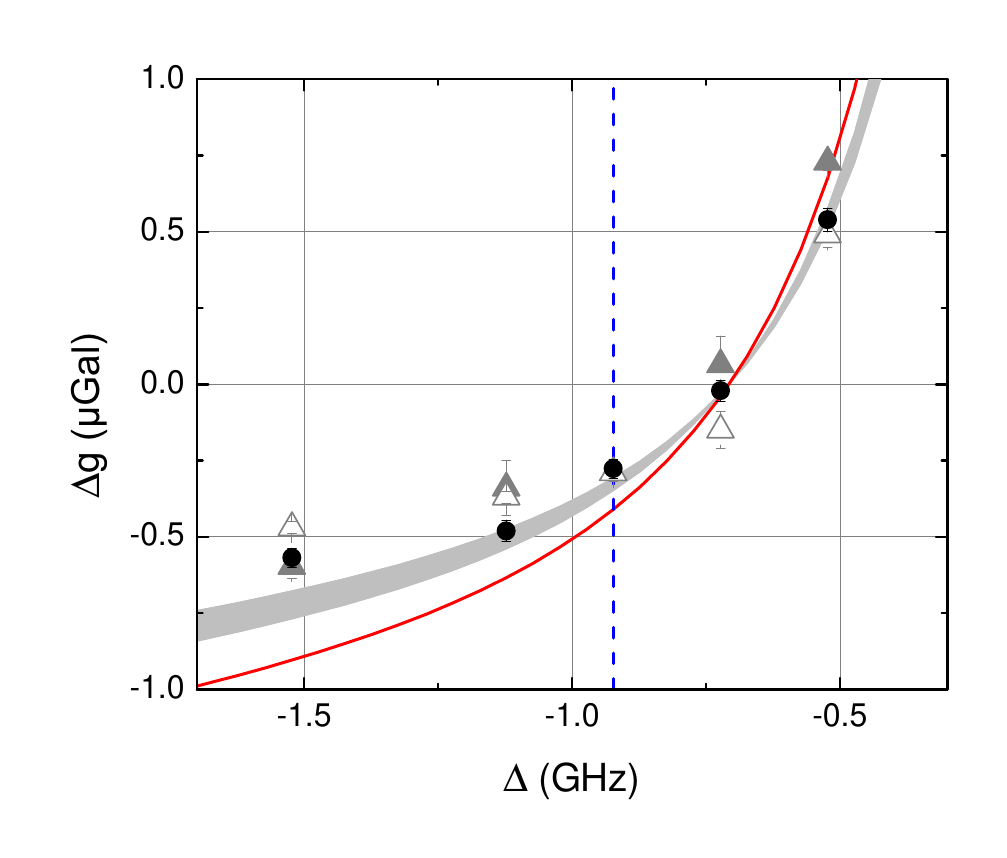}
\caption{(Color online) Effect of chirping only one of the two Raman lasers on the gravity measurement. Red line: calculation explained in
  section~\ref{section2}; filled gray area: calculation based on a Monte Carlo simulation of the interferometer explained in section~\ref{sec3}. Triangles: two sets of measurements at fixed Raman intensity, full (resp. open) triangles for full (resp. half) intensity; dots: measurements at a fixed Rabi frequency of 11.4$~$kHz. The -0.923$~$GHz Raman detuning used with the CAG is displayed with the dashed blue line. }
        \label{fig:lsvsdet}
\end{figure}

Yet, this shift depends on other parameters, such as $T$ and $\Delta$. The induced phase shift scales linearly with $T$, whereas the gravity phase shift scales quadratically. The influence of this effect on the gravity measurement thus reduces when increasing $T$. We plot as a solid red line the calculated bias on the gravity measurement $\Delta g$ as a function of the Raman detuning $\Delta$ in figure~\ref{fig:lsvsdet}. The effect is found to be monotonic with respect to the Raman detuning $\Delta$. Interestingly, operating at a detuning of $-707~$MHz nullifies the effect.

\section{Measurements}\label{sec3}

\subsection{Description in the experimental set-up}

In our Cold Atom Gravimeter (CAG) experiment, about $10^8$ $^{87}$Rb atoms are loaded within 80~ms in a three-dimensional magneto-optical trap (3D-MOT) from a 2D-MOT. After a far-detuned optical molasses phase, the lasers are adiabatically switched off within 200~$\mu$s and the atoms fall at a temperature of $2~\mu$K~\cite{Louchet-Chauvet2011}. During their free fall, an interrogation sequence takes place. In this study, this interrogation will be either a micro-wave spectroscopy (see section ~\ref{mw}) or a Raman interferometer (see section~\ref{intmeas}). Finally, after their interrogation, the atoms are detected at the bottom of the vacuum chamber thanks to a state selective detection method which allows to measure the populations of the atoms in each hyperfine state~\cite{LeGouet2008}. The total cycling time is 360~ms.

\subsection{Measurement of the DLS using microwave spectroscopy}\label{mw}

\begin{figure}
	\includegraphics{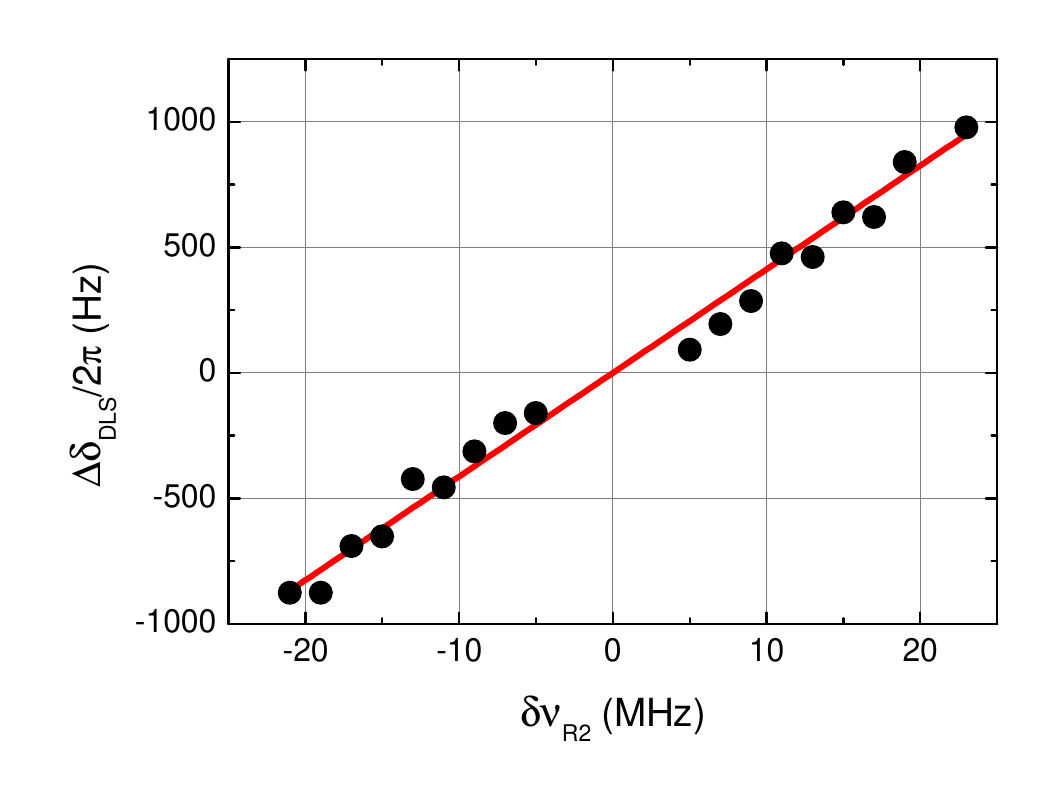}
	\caption{(Color online) Differential Light shift as a function of the change $\delta\nu_{R2}$ in the frequency of the second Raman laser ($R2$). Dots: measurements. Red continuous line: calculation.}
	\label{fig:lsvschirp}
\end{figure}

In a first series of measurements, we interrogate the atoms using microwave spectroscopy. At the end of the molasses, atoms are in the $|F=2\rangle$ state, populating all five $m_F$ sub-levels. 1~ms after releasing the atoms, we apply a vertical static magnetic field of 10~mG which lifts the degeneracy between the different magnetic sub-levels. 10~ms later, we switch on a 0.4~ms microwave pulse tuned close to the hyperfine transition in order to selectively address the $|F=2, m_F=0\rangle \rightarrow |F=1, m_F=0\rangle$ transition. By scanning the frequency of this microwave pulse across the hyperfine transition resonance, in the presence of the two Raman lasers set far detuned from the two photon transition, we measure the effect of the DLS as a shift $\delta_{DLS}$ of the resonance. We use this method to set the DLS
to zero adjusting the ratio between the two Raman lasers. Once this power ratio adjusted, we measure, using the same method, the change of the light shift as a function of the frequency of $R2$ ($\nu_{R2} = \nu_{R2}^{0} + \delta\nu_{R2}$), keeping the frequency of $R1$
fixed. Here, we deliberately change the frequency of only one laser (and not the frequencies of the two lasers by the same amount) in order to
emphasize the effect. Indeed, under this condition, we calculate using equations \ref{eq:lsf} and \ref{eq:lse} a change of $\Delta DLS$ of 41.3~Hz/MHz, about one order of magnitude larger than that while changing both laser frequencies (5~Hz/MHz) and thus easier to resolve. The
results of the measurements are displayed on figure~\ref{fig:lsvschirp}. We find a linear trend, of 41(1)~Hz/MHz, in perfect agreement with the
calculation, which validates our evaluation of the light shift for these lasers parameters.

 \subsection{Measurement of the effect of the DLS on the interferometer}\label{intmeas}

In a second series of measurements, we measure the effect on the DLS onto the phase of an atom interferometer. We now turn back the Raman lasers on resonance onto the two photon transition in order to drive the atom interferometer. After their release, atoms are selected in velocity and prepared in the $|F=1, m_F=0\rangle$ state using a sequence of microwave, pusher and Raman pulses. The Mach-Zehnder type interferometer realized with three Raman pulses ($\pi/2-\pi-\pi/2$) is performed in $2T=160$~ms. To keep the Raman lasers (of frequency $\omega_1$ and $\omega_2$, and wavevector $k_1$ and $k_2$) resonant with the atoms during their fall, the frequency of one of the two Raman lasers is chirped in order to compensate for the linearly increasing Doppler effect.

\begin{figure}[h]
	\centering
	\includegraphics[width=14 cm]{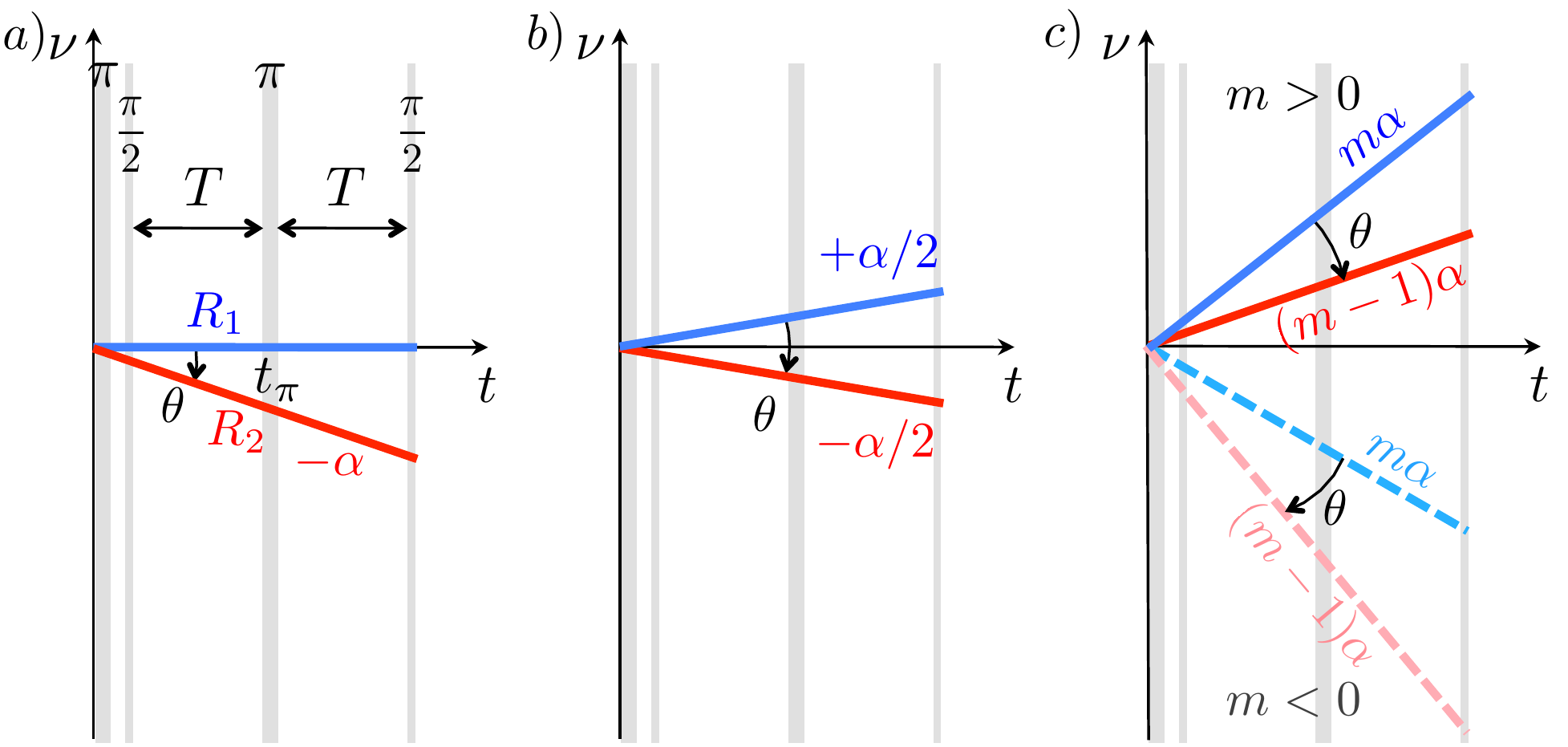}
	\caption{(Color online) Frequencies of the two Raman lasers $R1$ (in blue) and $R2$ (in red) during the interferometer for an effective wavevector pointing downwards.  The Raman $\pi$ velocity selection pulse and the $\pi/2-\pi-\pi/2$ interferometer pulses are represented in gray. Three cases are studied here: $a)$ the usual case used before this study: $R1$ is constant ($\alpha_1=0$) and only $R2$ is chirped at $\alpha_2=-\alpha$. $b)$ both $R1$ and $R2$ are chirped at $\alpha_1=+\alpha/2$ and $\alpha_2=-\alpha/2$ respectively. $c)$ $R1$ is chirped at $\alpha_1 = m\alpha$, and $R2$ is chirped at $\alpha_2=(m-1)\alpha$.  Here, $m$ is the magnification factor that ranges from $-4$ to $+4$.  }
	\label{fig:chirps}
\end{figure}

The total phase shift at the output of the interferometer is then given by $\Delta \Phi = \vec{k}_\mathrm{eff}\cdot \vec{g}~T^2-\alpha T^2$, where $\Delta \Phi$ is the phase difference between the two arms of the interferometer, $ \vec{k}_\mathrm{eff}$ the wavevector driving the atomic transitions ($\vec{k}_\mathrm{eff}~=~\vec{k}_{1}-\vec{k}_{2}$) and $\alpha$ the rate of the frequency chirp (Fig.~\ref{fig:chirps} $a$)).

Figure~\ref{fig:chirps} represents the frequencies of the two Raman lasers during the interferometer, in the case where the wavevector $\vec{k}_\mathrm{eff}$ is pointing downward ($k_\downarrow$). We now call $\alpha_i$ the frequency chirp of Raman laser $R_i$. In Case $a)$, the frequency of the first Raman laser $R1$ is kept fixed while the frequency of Raman laser $R2$ is chirped ($\alpha_1=0; \alpha_2=-\alpha$). This setting corresponds to the measurement procedure we have used so far. In Case $b)$, both Raman lasers are chirped in opposite   directions, with $\alpha_1=\alpha/2; \alpha_2=-\alpha/2$, so that their frequencies are fixed in the atomic frame. In that case, there is no change of the light shift, and thus no bias in the $g$ measurement. As the difference in the $g$ measurements between Cases $a)$ and $b)$ is expected to be small, and potentially difficult to resolve, we magnify the effect in Case $c)$, by increasing the frequency chirps applied to the lasers, while keeping their difference equal to $\alpha$. In practice, we apply chirp rate of $m\alpha$ to $R1$, and $(m-1)\alpha$ to $R2$, with $m$ an integer ranging in between $-4$ and $+4$. Then we performed differential gravity measurements between Cases $a)$ and $c)$ as a function of $m$. To remove most of the systematics effects, a $g$ measurement is obtained from the average of interleaved measurements performed with $\vec{k}_\mathrm{eff}$ oriented upward and downward~\cite{Louchet-Chauvet2011}. Such a $g$ measurement is in principle still biased by the Coriolis force, wavefront aberrations and the TPLS, but these biases are expected to be independent from the way the chirp rates are applied to the lasers. They are thus common to all cases and have no impact on the gravity differences between these cases.

\begin{figure}
	\includegraphics{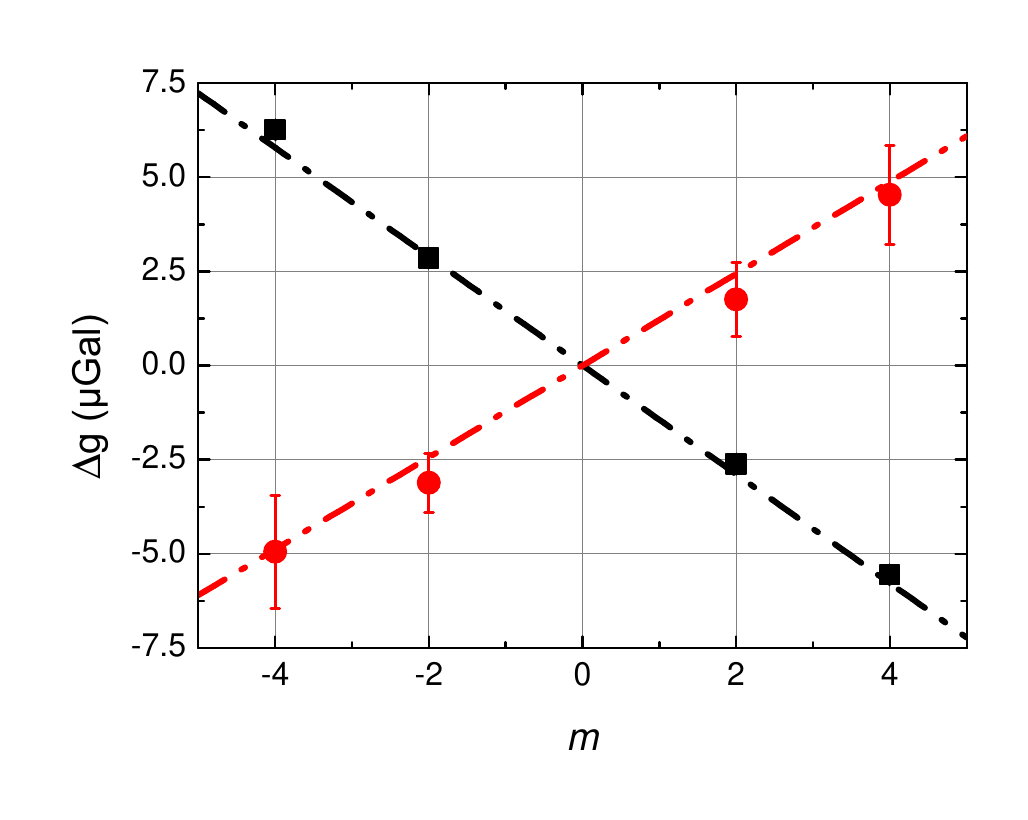}
	\caption{(Color online) Differential $g$ measurements, corresponding to the difference between Cases a) and c) of figure~\ref{fig:chirps}, as a function of the magnification $m$ of the chirp rate on the laser $R1$, ranging from $-4\alpha$ to $+4\alpha$.  Black squares: $\Delta$=-0.523~GHz, red dots: $\Delta$=-1.523~GHz. The lines are linear fit to the data with forced zero intercept for each data set.}
	\label{fig:dgvsrepchirp}
\end{figure}

Figure~\ref{fig:dgvsrepchirp} displays the results of the differential measurements between Cases $a)$ and $c)$, for different Raman detunings $\Delta$ of $-0.523~$GHz and $-1.523~$GHz at the full intensity of the Raman lasers. We find linear trends for each detuning with slopes of -1.45(3)~$\mu$Gal/$m$ at $-0.523~$GHz and 1.22(19)~$\mu$Gal/$m$ at $-1.523~$GHz. Multiplying these slopes by $m~=~$-0.5 gives the difference between Cases $a)$ and $b)$, and thus the expected bias on the gravity measurement when operating the gravimeter with the settings of Case $a)$. We repeat the above measurement to determine the bias at other Raman detunings keeping the Raman intensity ($I$) at maximum and display the results on figure~\ref{fig:lsvsdet} as full triangles. As the Rabi frequency scales as $I/\Delta$, measurements at different detunings correspond to different Rabi frequencies and thus different Raman pulse durations. Using the same procedure, we also performed measurements at half Raman laser intensity, displayed as opened triangles, and finally at a fixed Rabi frequency of 11.4$~$kHz, displayed as dots in figure~\ref{fig:lsvsdet}. In the latter configuration, we vary the intensity of the Raman lasers with the detuning in order to keep the Rabi frequency and thus the pulse durations constant. These three series give similar results, which confirm that the effect is independent of the Rabi frequency. When compared with the result of the calculation of section~\ref{section2}, we obtain the correct trend, though the measured effect is weaker. For a better comparison we performed a Monte Carlo simulation of the interferometer~\cite{Gillot2015}, taking into account our experimental parameters. The atoms are drawn randomly in a Gaussian position distribution of $\sigma~=~0.5~$mm and in a Lorentzian $b$ velocity distribution~\cite{Farah2014, Sortais2000} ($f(v)=A/(1+(v-v_0)^2/v_c^2)^b$), with parameters $v_c~=~16.5~$mm and $b~=~$2.4. The Raman beam is modeled with a Gaussian beam of 12$~$mm waist ($1/e^2$ radius). The simulation includes the effect of the Raman velocity selection and the response of the detection~\cite{Louchet-Chauvet2011, Gillot2015}. It calculates the evolution of the atomic state in the interferometer, taking into account the trajectory of the atoms in the laser beams (and thus the coupling inhomogeneities), the Raman light shifts, as well as other effects which are not relevant here (such as the Coriolis acceleration and the TPLS). The result of the simulation is displayed as a gray filled area as it takes into account the uncertainty in the experimental parameters. In particular, we estimate that the adjustment of the Raman pulse parameters can deviate from optimal settings ($\Omega\tau_{\pi/2}~=~\pi/2$) by 5$~\%$ at most from measurement to measurement. The agreement with the measurements is better, though we find resolved differences between measurements and calculations of up to 0.2~$\mu$Gal for larger values of the detuning.

\section{Finite speed of light}

The propagation delay due to the finite speed of light introduces a bias on the $g$ measurement in classical free-fall-corner-cube
gravimeters~\cite{Kuroda1991}. Similar effects have been put in evidence in atom gravimeters: a significant bias of up to 10.4$~\mu$Gal was found in \cite{Peters2001} when dropping atoms and performing the measurement for a single orientation of the Raman wavevector.

To calculate this effect, we express the total interferometer phase shift $\Delta\Phi_{tot}$ following~\cite{Borde2001}:
\begin{equation}
\Delta\Phi_{tot}=\phi(-T,\vec{z})-2\phi(0,\vec{z})+\phi(T,\vec{z})\nonumber 
\end{equation}
with the Raman laser phase difference $\phi$ imprinted on the atomic phase at the three pulses $\pi/2-\pi-\pi/2$, considering that the $\pi$ pulse occurs at $t=t_\pi=0$. 

Taking into account the delay due to the propagation of the light from the very position where the phase difference between the lasers is measured (and phase locked) to the position of the atoms, we find, after some algebra,

\begin{equation}
\Delta\Phi_{tot}=(\alpha_1-\alpha_2)T^2-(\omega_{1,0}\widehat{k_1}-\omega_{2,0}\widehat{k_2})\cdot\frac{\vec{g}}{c}T^2-2(\alpha_1\widehat{k_1}-\alpha_2\widehat{k_2})\frac{\vec{v}_0}{c}T^2,
\label{eqDPHItot}	
\end{equation}
with $\omega_{1,0}$, $\omega_{2,0}$ and $\vec{v}_0$ the Raman laser frequencies and the atom velocity at $t=0$; $\widehat{k_1}$ and
$\widehat{k_2}$ are unit vectors along the direction of propagation of the lasers. By determining the position of the dark fringe, for which
$\Delta\Phi_{tot}=0$, we extract from equation~\ref{eqDPHItot} the value of $g$,
\begin{eqnarray}
g&=&\frac{\alpha_1-\alpha_2}{k_\mathrm{eff}}+2\frac{v_0}{c}\frac{\alpha_1+\alpha_2}{k_\mathrm{eff}},
\label{eqg}	
\end{eqnarray}
with $\vec{k}_\mathrm{eff}~=~(\omega_{1,0}\widehat{k_1}-\omega_{2,0}\widehat{k_2})/c$ the effective wave vector at the $\pi$ pulse.

Introducing $g_0=(\alpha_1-\alpha_2)/k_\mathrm{eff}$, equation~\ref{eqg} gives:
\begin{equation}
g=g_0\Bigg(1+2\frac{v_0}{c}\frac{\alpha_1+\alpha_2}{\alpha_1-\alpha_2}\Bigg),
\end{equation}
which is identical to the equation 48 of \cite{Peters2001}. This equation shows that if the Raman lasers are not chirped symmetrically, the value of $g$ deviates from $g_0$. In our case, where we usually operate according to case $a)$ of figure~\ref{fig:chirps}, we find a correction that amounts to $6.15~\mu$Gal for a single $\vec{k}_\mathrm{eff}$ measurement, as the velocity at the $\pi$ pulse is $v_0=0.94~$m/s.

Magnifying the chirp rate of the $R1$ and $R2$ frequency by $m$, we expect a linear scaling of this correction as a function of $m$ according to
\begin{equation}
\Delta g=4 m g_0 v_0/c.
\label{eq1}
\end{equation}
This gives a correction of $(12.29~\times~m)~\mu$Gal.

In fact, the sign of this correction depends on the orientation of $\vec{k}_\mathrm{eff}$. When alternating the $k_{\uparrow}$ and
$k_{\downarrow}$ configurations, this effect is removed when calculating the averaged value $g_{mean}=\frac{1}{2}\sum_g=\frac{1}{2}(g_\uparrow+g_\downarrow)$, and thus does not contribute to the results of the previous section. Alternatively, this correction can be obtained by calculating half the difference ($\frac{1}{2}\Delta_g$). To separate this contribution from other systematic effects that affect this difference, we perform differential measurement taking the $m=0$ case as a reference according to
\begin{equation}
	\Delta g_{eq}=\frac{1}{2}(g_\uparrow-g_\downarrow)_{m=0}-\frac{1}{2}(g_\uparrow-g_\downarrow)_{m}.
\label{eqDgeq}
\end{equation} 

\begin{figure}
	\includegraphics{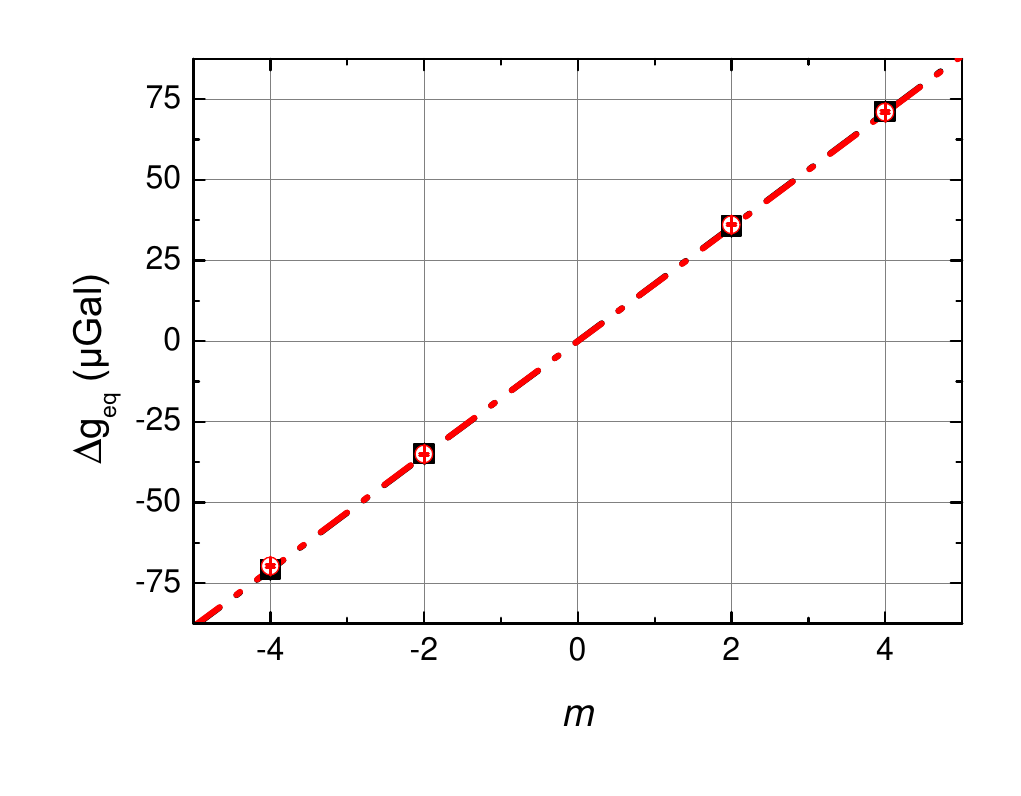}
	\caption{(Color online) Measurement of $\Delta g_{eq}$ according to equation~\ref{eqDgeq}, as a function of the magnification $m$ of the
          chirp rate on the laser $R1$, ranging from $-4\alpha$ to $+4\alpha$. Black squares: $\Delta$=-0.523~GHz, red dots: $\Delta$=-1.523~GHz. The (nearly overlapping) lines are a linear fit to the data with forced zero intercept for each data set.}
	\label{fig:halfsum}
\end{figure}

Results are displayed on figure~\ref{fig:halfsum} for two different Raman detunings $\Delta$=-0.523~GHz and $\Delta$=-1.523~GHz. The figure shows, as expected, no dependence in the detuning $\Delta$, and linear behaviors of 17.73(4)~$\mu$Gal for $\Delta$=-0.523~GHz and 17.68(6)~$\mu$Gal for $\Delta$=-1.523~GHz respectively.

These slopes differ from the result of equation~\ref{eq1}. This arises from the fact that, in the experiment, the effective wavevector at the $\pi$ pulse varies with $m$, contrary to the calculation leading to equation~\ref{eq1}. This leads to an additional contribution given by
\begin{equation}
\Delta g=2 m g t_{\pi}/c,
\label{eq2}
\end{equation}
where $t_{\pi}$ is the duration between the start of the chirp and the moment of the $\pi$ pulse. It amounts to $(5.50~\times~m)~\mu$Gal. Adding equations \ref{eq1} and \ref{eq2}, we find $17.79~\mu$Gal in very good agreement with the measured slopes. This confirms our expectation that alternating between two opposite directions of the Raman wavevector $\vec{k}_\mathrm{eff}$ allows to separate the effect of finite speed of light from that related to the change of the light shift induced by the chirp of the Raman lasers.  Thus, the measurements of   section~\ref{sec3} is not affected by the finite speed of light.  Moreover, it strengthens our claim that our $k$-reversal protocol for measuring $g$ rejects the effect of the finite speed of light, as stated in the accuracy budgets of the CAG when participating to metrological international comparisons~\cite{Merlet2010, Louchet-Chauvet2011b, Jiang2012, Francis2013, Gillot2014, Francis2015}.

\section{Conclusion}

We have shown that chirping only one of the two Raman lasers to compensate for the change in the Doppler shift in a Raman interferometer based gravimeter induces a time dependent light shift, that leads to a phase shift. As this phase shift depends on the direction of the Raman wavevector, it leads to a bias in the gravity measurement that is not suppressed by the $k$-reversal technique. We have measured the amplitude of this bias for our typical parameters, that amounts to $0.30(2)~\mu$Gal. This bias, which is smaller than our current claimed accuracy of $4.3~\mu$Gal, was not accounted for in our previous measurements, and in particular, during the international comparison campaigns in which we participated.

We have performed measurements of this effect as a function of the Raman detuning, and show that this shift is independent of the Rabi frequency and varies from $0.6~\mu$Gal to $-0.5~\mu$Gal when the detuning varies from $-0.5~$GHz to $-1.5~$GHz. 

This effect decreases for increasing interferometer durations and thus is lower for the atomic fountain geometries such as in~\cite{Peters2001, Zhou2011, Hu2013, Hauth2013} than when the atoms are simply dropped~\cite{Bodart2010, Bidel2013, Wu2014}. In any case, it can be suppressed by chirping both Raman lasers in opposite directions. In this case, the frequencies of the lasers are constant in the frame of the atoms during their fall, so that the light shift, if any, is fixed, and rejected by the symmetry of the interferometer.

We have also performed a study of the influence of the finite speed of light which is the subject of recent controversy in the context of free falling corner cube gravimeters~\cite{Nagornyi2011, Rothleitner2011, Nagornyi2011b, Rothleitner2011b, Rothleitner2014, Baumann2015}. Our measurements confirm the validity of the analysis of~\cite{Peters2001} for the case of an atom gravimeter.

\begin{acknowledgments}

B.C. thanks the Labex FIRST-TF for financial support. Authors thanks A. Landragin for useful discussions and B. Fang for carefully reading and correcting the manuscript. This research is carried on within the kNOW project, which acknowledges the financial support of the EMRP. The EMRP was jointly funded by the European Metrology Research Programme (EMRP) participating countries within the European Association of National Metrology Institutes (EURAMET) and the European Union. 

\end{acknowledgments}


\end{document}